# Disorder-induced modal alignment in transmission eigenchannels enables control of wave transport

Israel Kurtz[1,2], Yiming Huang[1,2,3], Zhou Shi[1,2,4], and Azriel Z. Genack[1,2]

[1]*Department of Physics, Queens College of the City University of New York, Flushing, New York 11367, USA*

[2]*Physics Program, The Graduate Center of the City University of New York, New York, New York 10016, USA*

[3]*Jinhua No.1 High School, Zhejiang, 321000, China*

[4]*Lightera Labs, 19 School House Road, Somerset, New Jersey 08873, USA*

The transmission eigenchannels of a disordered medium range from unity transmission and high internal energy density to vanishing transmission and low internal energy density. Although the transmission matrix underlies central ideas in mesoscopic electronic transport, it cannot be measured directly in electronic systems. For classical waves, however, the measurable matrix gives access to eigenchannels that control transmission and energy density within the medium. Coherence underlies transmission-eigenvalue scaling, the bimodal distribution in diffusive samples, and Anderson localization. What has remained unresolved is how interference among modal contributions is organized within individual eigenchannels throughout the sample. Here, microwave transmission-matrix measurements resolve the complex contribution of each incident mode to every transmitted mode of each eigenchannel. In simulations, we construct a flux matrix that generalizes the transmission matrix to every depth and further separate each internal modal contribution into forward- and backward-propagating components. For each eigenchannel, the directional flux in every mode throughout the sample factorizes into the squared modal amplitudes of the incident eigenchannel, coupling between waveguide modes, and modal alignment. In high-transmission eigenchannels, the contributions align increasingly constructively with depth, approaching nearly perfect alignment in the highest channel near the localization crossover. In low-transmission eigenchannels, contributions interfere destructively, producing vanishing transmission at a transmission zero. Because the contributions remain appreciable as their coherent sum approaches zero, transmission far below the noise floor of conventional transmission measurements can be determined. In our measurements, the lowest transmission eigenvalue near a transmission zero is more than nine orders of magnitude below the highest. These results provide a depth-resolved decomposition of the modal contributions and interference among them, which produces the broad range of eigenchannel transmissions.

## 1. INTRODUCTION

When a classical wave passes through a disordered medium, multiple scattering transforms the wave into a stationary speckle pattern of flux. A framework for understanding and controlling transmission is provided by the transmission matrix, whose elements are ratios of outgoing to incident complex flux amplitudes [1–9]. Although the magnitudes of the elements of the transmission matrix are comparable across eigenchannels, their eigenvalues span the range from complete [10–12] to vanishing transmission [13–15], with striking variations in energy density throughout the sample [16–20]. These phenomena are consequences of phase coherence in scattering media: interference governs the scaling [1,21] and statistics [22] of transmission eigenvalues, including the bimodal distribution in diffusive systems [1–3,5,6] and Anderson localization [23]. What has not been resolved is the internal modal construction of a transmission eigenchannel, how the flux amplitude in each waveguide mode is formed from contributions of all incident modes, how their interference is organized throughout the sample, and how it combines with other factors to determine eigenchannel structure.

The transmission matrix, first developed to describe electronic conductance in mesoscopic samples in which the wavefunctions of electronic quasiparticles remain temporally coherent, provides a complete framework for the scaling and statistics of electronic transport [1–3,5,6,24–28]. In practice, however, the transmission matrix cannot be measured directly in electronic systems because conductance is measured between thermal reservoirs rather than between well-defined coherent modes. The underlying phase coherence in electronic transport can therefore only be inferred from measurements of conductance fluctuations and quantization in a single mesoscopic sample [28–32].

For classical waves, by contrast, the transmission matrix can be measured directly [7–10,24,33–36], allowing the transmission eigenvalues and eigenchannels to be determined. In two-dimensional samples, the internal flux matrix has given the energy deposition eigenvalues [19]. Coherent control of the incident fields had earlier enabled focusing through [7,37–39] and imaging in [40–42] random media. Direct measurement of the transmission matrix [8] subsequently enabled transmission eigenvalues to be measured over a high dynamic range [16,17,33] and a variety of transmitted waveforms to be excited [36]. Here, we determine the relations among the complex modal flux amplitudes forming each eigenchannel, identify the factors governing the eigenchannel flux, and track their evolution through the medium.

As first recognized by Landauer [25], quantum transport can be characterized by its classical analog, the transmittance, $T$, which is equivalent to the dimensionless conductance $G/(e^2/h)$ [25–27]. Here, $G$ is the electronic conductance, $e^2/h$ is the quantum of conductance, and $T$ is the sum of flux transmission coefficients between the $N$ incident and outgoing propagating waveguide modes. The ensemble average of the dimensionless conductance is denoted by $g$, $g = \langle T \rangle$. In the waveguide or wire geometries, $T = \sum_{s,m}^{N} |t_{ms}|^2 = \mathrm{Tr}(tt^\dagger)$, where the $t_{ms}$ are the elements of the transmission matrix [27]. Equivalently, $T$ can be expressed as the sum of the $N$ eigenvalues of $tt^\dagger$, $T = \sum_{n=1}^{N} \tau_n$. Dorokhov showed that each transmission eigenvalue scales with a distinct localization length. This leads to the bimodal distribution of transmission eigenvalues for diffusive waves, in which roughly $g$ eigenvalues exceed $1/e$ and carry most of the conductance, whereas the remaining eigenchannels transmit only weakly [2,3,6,43,44]. These results establish the scaling and statistics of transmission eigenvalues but do not resolve the modal composition and interference within individual eigenchannels.

In addition to transmission eigenvalues, transmission eigenchannels possess distinct average velocities, $\mathrm{v}_n$, which link the flux, $\tau_n$, to the output linear energy density, $u_n(L)$, through the relation $\tau_n = u_n(L)\mathrm{v}_n$ [45]. The eigenchannel velocities $\mathrm{v}_n$ are defined in the Supplemental

Material in terms of the weighted velocities $v_m$ of the modes of the empty waveguide [46]. The eigenchannel velocities have been characterized statistically at the input and output surfaces [45], but their behavior within the sample has not been previously reported.

Using microwave measurements, we determine, at the output, the alignment of the complex contributions from the incident waveguide modes composing each transmission eigenchannel to each transmitted waveguide mode. Complementary numerical simulations resolve the field within the sample into forward- and backward-propagating components, allowing us to track the spatial evolution of the flux for all eigenchannels. We show that the flux amplitude at any depth, including the output where it gives the transmission eigenvalue, is determined by three factors: the incident modal profile, the coupling of incident modes to modes within the sample, and the collective alignment of the corresponding complex modal contributions, which we term modal alignment. Modal alignment varies systematically with depth to produce constructive interference in eigenchannels whose transmission exceeds the ensemble average $g/N$ and destructive interference in those with transmission below $g/N$.

## 2. METHODS

### A. Experimental measurement of the transmission matrix

We measured the transmission matrix of a disordered medium in a copper waveguide supporting $N = 62$ propagating modes. For this sample, the dimensionless conductance was $g = 3.2$. The sample, of length $L = 40$ cm, consisted of alumina spheres of diameter 0.95 cm and refractive index 3.14, each embedded in a Styrofoam shell and randomly packed in the waveguide, giving an alumina volume filling fraction of 0.069. Field transmission coefficients between source and receiver antennas were obtained using a vector network analyzer. Wire antennas, formed by extensions of the central conductors of coaxial cables, were scanned over grids of input and output positions, with measurements at each position for two orthogonal antenna orientations (Fig. 1(a)). Rotating the tube about its axis generated new random configurations. Absorption was weak and did not alter the observed trends; additional experimental details and analysis of its impact are given in the Supplemental Material [46].

**B. Poles and zeros of the transmission matrix**

Wave propagation through a disordered medium can be described from complementary perspectives: as a superposition of quasinormal modes or in terms of the transmission matrix. These descriptions are linked through the analytic structure of the determinant of the transmission matrix, $\det(t)$. The field within the sample is the coherent sum of contributions from the quasi-normal modes excited by the incident radiation [47–50], while $\det(t)$ can be factorized in terms of its zeros $\{\eta_i\}$ and poles $\{\Omega_n\}$ [13],

$$\det(t) \propto \frac{\prod_i(\omega - \eta_i)}{\prod_n(\omega - \Omega_n)}. \tag{1}$$

The poles $\Omega_n = \omega_n - i\gamma_n$, where $\gamma_n$ is the halfwidth of the $n$th mode, are the complex resonance frequencies of the quasi-normal modes. These lower-half-plane poles coincide with those of the scattering matrix and give the complex frequencies of the resonant states of the medium. The zeros are frequencies at which $\det(t)$ vanishes because of destructive interference among the modal contributions to transmission.

Both poles and zeros appear as singularities in the complex-frequency map of the phase, $\arg\{\det(t)\}$. The phase of $\det(t)$ increases (decreases) by $2\pi$ upon encircling a transmission zero (pole) in the counterclockwise direction. Figure 6(a) shows a phase map in the complex plane, in which poles and zeros appear as phase singularities.

In a lossless system, $\det(t)$ is real for real $\omega$, which implies that its zeros are symmetric with respect to the real axis and are therefore either real or part of a complex-conjugate pair, $\eta_i = \omega_i \pm i\gamma_i$ [13]. The lowest transmission eigenvalue vanishes only when the frequency coincides with a real-axis zero of $\det(t)$. For a complex-conjugate pair of zeros off the real-frequency axis, vanishing transmission requires gain or loss to shift one zero onto the real axis. Uniform loss or gain shifts the phase map downward or upward, respectively, by $\gamma$, where $\gamma$ is the field-decay rate for loss and the field-growth rate for gain.

### C. Flux matrix and depth-resolved decomposition

The flux matrix, $t_{ms}(z)$, is the extension of the transmission matrix formalism to all sample depths $z$. It gives the ratio of the sum of forward- and backward-propagating flux components in a mode $m$ inside the sample to the flux in incoming mode $s$. Further details are given in the Supplemental Material [46]. We decompose the flux matrix into forward- and backward-propagating components,

$$t(z) = t^{+}(z) + t^{-}(z), \tag{2}$$

where $t^{+}(z)$ and $t^{-}(z)$ correspond to forward- and backward-propagating components, respectively. This decomposition is obtained from fields computed using recursive Green's function simulations [51,52]; details are given in the Supplemental Material [46]. The separation into forward and backward contributions enables computation of the net flux in each eigenchannel, which is constant throughout the sample and equal to the transmission eigenvalue. At $z = L$, $t(z)$ reduces to the transmission matrix, $t(L) = t$.

A mathematically equivalent quantity, termed the deposition matrix, was previously introduced and measured in two-dimensional disordered optical waveguides open to the third dimension [19,20]. There, the singular value decomposition (SVD) of the matrix at each depth was used to determine the incident wavefront that maximizes the energy deposited at a chosen cross section. Here, instead, we use the flux matrix $t(z)$ to follow the flux generated throughout the sample by the transmission eigenchannels of the full transmission matrix. This formulation provides the basis for the results that follow, where we show how interference among the waveguide modes determines propagation within each eigenchannel.

**D. Numerical simulations**

Numerical simulations are performed for quasi-one-dimensional disordered waveguides using recursive Green's function methods [51,52]. The medium is discretized on a square lattice with lattice constant $a = \frac{\lambda_{\min}}{2\pi}$, where $\lambda_{\min}$ is the minimum wavelength present in the simulation. Disorder is introduced through random variations of the dielectric constant within a range $\Delta\varepsilon$ about unity. Additional details of the numerical implementation are given in the Supplemental Material [46].

## 3. RESULTS AND DISCUSSION

### A. Experimental measurements of transmission eigenvalues

We measured the transmission matrix of a cylindrical copper waveguide randomly packed with dielectric spheres. The experimental apparatus used is shown schematically in Fig. 1(a); the electric-field transmission coefficients between antennas on the input and output faces of the waveguide were measured using a vector network analyzer. The antennas are positioned in the input and output surface planes and are rotated by 90° to obtain four orthogonal field-transmission coefficients.

Transmission eigenchannels are obtained from the SVD of the transmission matrix $t$, expressed in the basis of the propagating waveguide modes of the empty waveguide, $t = U\Lambda V^{\dagger}$, where $V$ and $U$ are unitary matrices with elements $v_{sn}$ and $u_{mn}$, and $\Lambda$ is a diagonal matrix with elements $\lambda_n = \sqrt{\tau_n}$. For unit incident flux in the $n$th eigenchannel, the amplitudes of the $s$th and $m$th waveguide modes at the input and output faces are $v_{sn}$ and $\lambda_n u_{mn}$, respectively [6,9]. Modes are indexed by descending group velocity $\mathrm{v}_m$, and transmission eigenchannels in order of descending transmission $\tau_n$.

Spectra of the transmission eigenvalues $\tau_n$ of the highest- and lowest-transmission eigenchannels in a sample of length $L = 40$ cm are shown in Fig. 1(b). The measured spectra of the highest and lowest transmission eigenvalues, $\tau_1$and $\tau_{62}$, for this configuration show that $\tau_{62}$ is approximately nine orders of magnitude smaller than $\tau_1$. The value of $\tau_{62}$ lies seven orders of magnitude below the noise floor of the transmission measurement, which is set primarily by thermal drift during the 40-hour acquisition time [14].

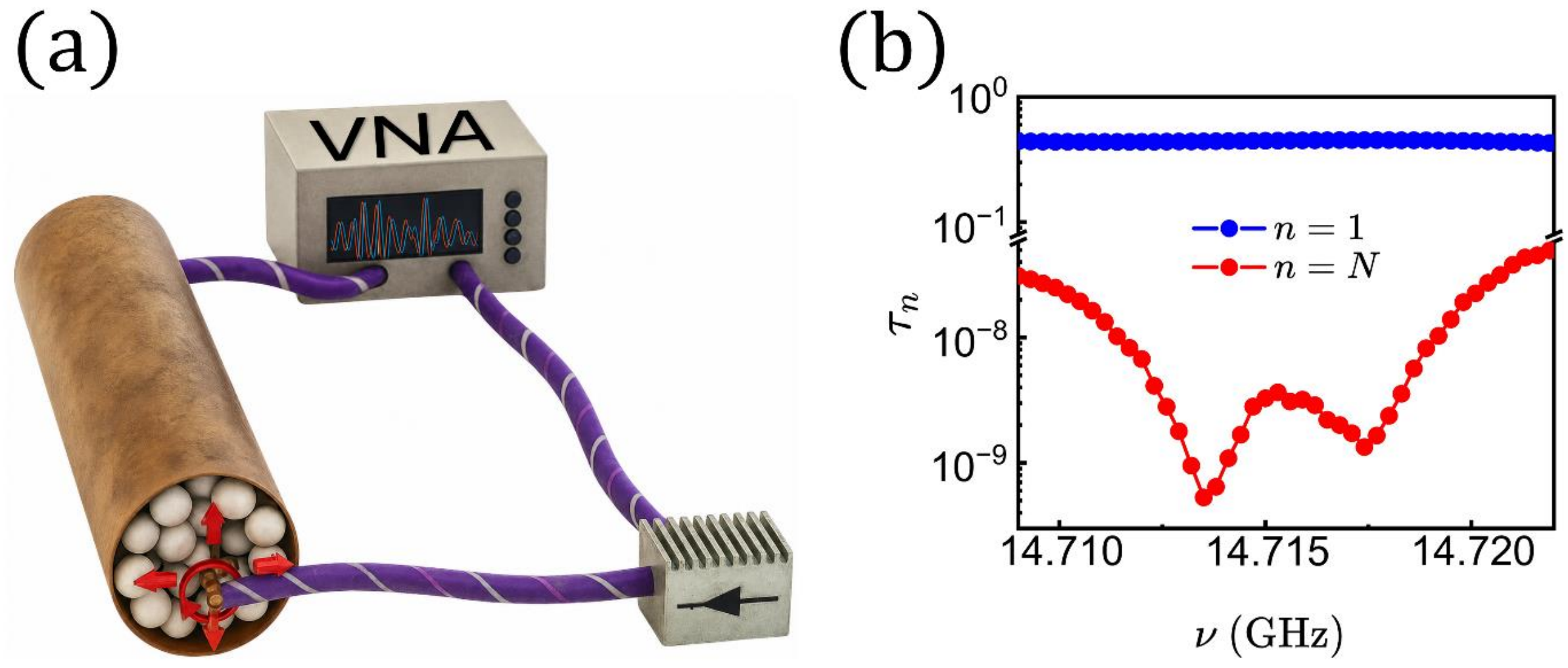

FIG. 1. Measurements of the transmission matrix. (a) Schematic of the experimental setup. Field transmission coefficients are measured using a vector network analyzer (VNA). The antennas are scanned over a grid of points in two orthogonal orientations on the input and output surfaces of a cylindrical copper waveguide randomly packed with dielectric spheres. (b) The spectrum of transmission eigenvalues for $n = 1$ and $n = N = 62$ in a single configuration. The measured dynamic range spans nine orders of magnitude, from the highest transmission eigenvalue to the lowest at transmission-zero frequencies $\nu_{TZ}$.

### B. Modal decomposition of transmission eigenchannels

To resolve the modal contributions and their interference, which underlie the broad range of transmission eigenvalues, we express the flux amplitudes of each transmission eigenchannel in the basis of the propagating modes of the empty waveguide. Unlike the quasinormal modes of the disordered medium, which are sample-specific and not known *a priori*, the waveguide modes have fixed and known spatial profiles. The waveguide-mode basis therefore provides a natural framework for analyzing the impact of interference within transmission eigenchannels.

The flux amplitude of the $m$th waveguide mode of the $n$th transmission eigenchannel is the coherent sum of contributions from all incident waveguide modes, $f_{mn}(z) = \sum_{s=1}^{N} t_{ms}(z)v_{sn}$. Here, $t_{ms}(z)$ is the $m, s$ element of the flux matrix at depth $z$ with its forward- and backward-propagating components given by $t_{ms}^{+}(z)$ and $t_{ms(}^{-}z)$, respectively. Interference among the incident-mode contributions determines the flux in each waveguide mode, while the fluxes of the orthogonal modes add incoherently. At the output, $\tau_n = \sum_{m=1}^{N} |f_{mn}|^2$. This decomposition of each modal flux amplitude into contributions from the incident waveguide modes allows us to track their amplitudes and phases with depth.

## C. Modal alignment at the output

The role of modal alignment in producing the diversity of transmission eigenvalues is illustrated experimentally in Fig. 2 for a sample with dimensionless conductance $g = 3.2$. Figure 2(a) shows constructive interference in the formation of the $m = 1$ output mode of the $n = 1$ eigenchannel, $f_{11}(L)$, at a single frequency through the alignment of complex vectors (phasors) representing contributions from the incident modes. We quantify the alignment of the contributions to the $m$th waveguide mode of the $n$th eigenchannel by the modal-alignment parameter

$$\kappa_{mn} = \frac{|\sum_s t_{ms} v_{sn}|^2}{\sum_s |t_{ms} v_{sn}|^2}. \quad (3)$$

Thus, $\kappa_{mn}$is the coherent modal flux normalized by the corresponding incoherent sum; values above or below unity indicate constructive or destructive alignment, respectively. For the example in Fig. 2(a), $\kappa_{11} = 4.24$. Its average over the measured configurations is also 4.24, compared with unity for random alignment, demonstrating strong constructive alignment in the highest-transmission eigenchannel.

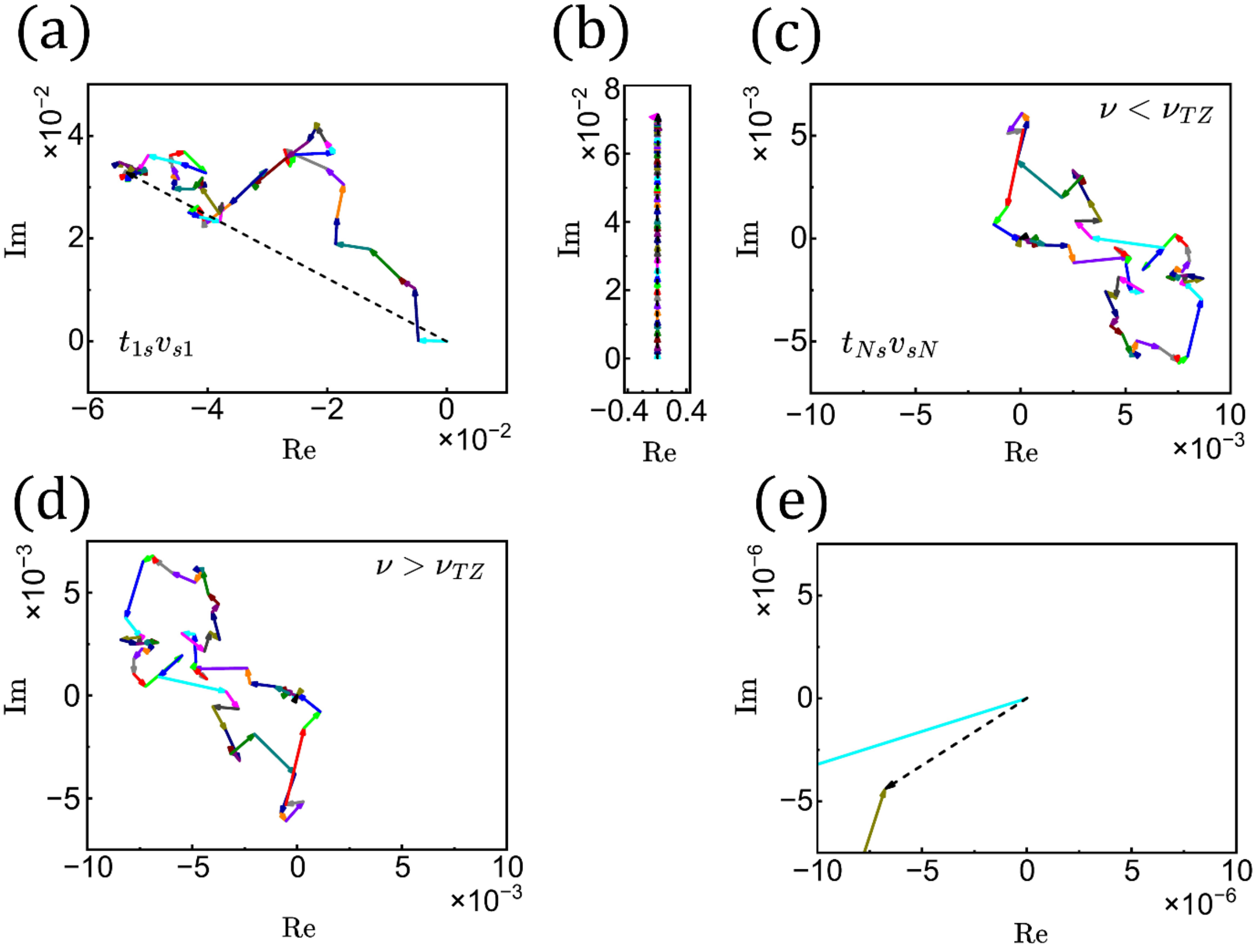

FIG. 2. Measurements of constructive and destructive interference in high- and low-transmission eigenchannels. (a) Vector diagram of the $N = 62$ complex flux components, $t_{1s}v_{s1}$ (colored arrows), and their resultant, $f_{11} = \sum_{s=1}^{62} t_{1s}\, v_{s1}$ (dashed black arrow), which gives the flux amplitude in the m = 1 mode of the n = 1 transmission eigenchannel at $\nu = 14.7168$ GHz. (b) Flux components $t_{1s}v_{s1}$, averaged over 801 frequencies in the range $14.7 - 14.94$ GHz for three random samples of length $L = 40$ cm. The resultant for each sample is aligned along the positive y-axis. (c)-(e) Vector sums $f_{62,62} = \sum_{s=1}^{62} t_{62s}\, v_{s62}$ at frequencies just below ($\nu = 14.7135 < \nu_{TZ}$) (c) and just above ($\nu = 14.7138 > \nu_{TZ}$) (d) the transmission zero $\nu_{TZ}$. The 62 contributions $t_{62s} v_{s62}$ interfere destructively, so the resultant is too small to be visible in (c) and appears only in (e) when the scale is expanded by a factor of $10^3$. The average phase shift of the 62 phasors across the transmission zero from (c) to (d) is $175° \pm 4°$.

By contrast, the lowest-transmission eigenchannel exhibits nearly complete destructive interference near a transmission zero [10]. The 62 complex phasor contributions $t_{62s} v_{s62}$, whose coherent sum gives the resultant $f_{62,62}$, are shown at frequencies just below and above the transmission zero in Figs. 2(c) and 2(d). At these frequencies, $\kappa_{62,62}$ equals $6.6 \times 10^{-7}$ and $7.8 \times 10^{-7}$, respectively. The destructive interference renders the resultant too small to be visible at the scale of Figs. 2(c) and 2(d); it becomes apparent only when the plot is magnified by a factor of $10^3$ in Fig. 2(e). Near-complete destructive interference is observed in all modes of the lowest-transmission eigenchannel.

Although the resultant $f_{62,62}$, formed by the coherent sum of the 62 contributions $t_{62s} v_{s62}$, is extremely small, the individual phasor amplitudes remain appreciable: their average magnitudes $|t_{62s} v_{s62}|$ in Figs. 2(c) and 2(d) are only a factor of 3 smaller than the corresponding magnitudes $|t_{1s} v_{s1}|$ in Fig. 2(a). Thus, the ultrasmall values of $\tau_{62}$ arise from near-perfect destructive interference rather than from small individual phasor amplitudes. The minima in the spectrum of $\tau_{62}$in Fig. 1(b) are determined by the proximity of the nearest measurement frequency to the transmission-zero frequency, $\nu_{\mathrm{TZ}}$ (see Supplemental Material [46]).

The 62 phasors undergo an average phase shift of $175° \pm 4°$across the transmission zero [Figs. 2(c) and 2(d)]. In simulations, the phase shift across a real-axis transmission zero approaches $180°$as the frequency step decreases.

### D. Evolution of modal alignment with sample depth

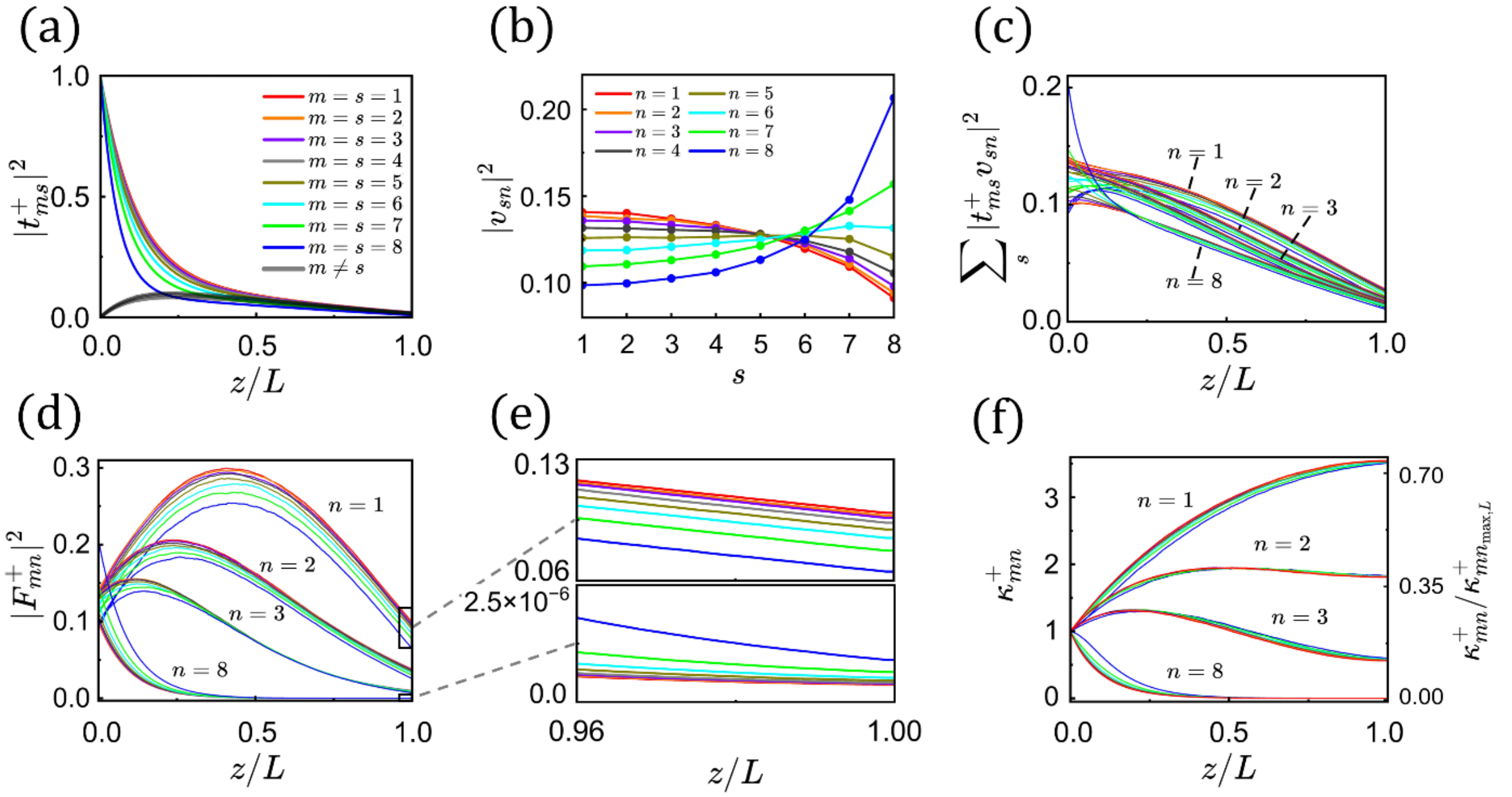

FIG. 3**.** Profiles of modal characteristics that determine the modal flux in transmission eigenchannels of random media. Simulations are carried out for an ensemble of $7.15 \times 10^5$ random samples with $W = 26a$, $L = 1000a$, $\Delta\varepsilon = 0.3$, $\ell = 133a$, and $g = 1.1$ at a frequency supporting $N = 8$ channels. (a) As the wave is scattered in the medium, the forward-propagating flux in input modes $s$ is redistributed among internal modes $m$. (b) The weights of modes with lower velocities increase for transmission eigenchannels with higher indices $n$. (c) The incoherent sum of the square of flux amplitudes seeded by incident modes decreases both with group velocity $\mathrm{v}_m$ and eigenchannel index $n$. (d) The forward flux, given by the coherent sum $|f^+_{mn}|^2 = |\sum_{s=1}^{N=8} t^+_{ms} v_{sn}|^2$ varies strongly with eigenchannel index $n$. (e) For high-transmission

eigenchannels, higher-velocity modes contribute more strongly to the flux, whereas this trend reverses for lower-transmission eigenchannels. (f) The degree of modal alignment among contributions to the forward-propagating modal flux $\kappa_{mn}^{+}$ quantifies the ratio of coherent (d) to incoherent flux (c).

We now use simulations to examine how the broad range of modal fluxes across transmission eigenchannels develops with depth. We focus here on the forward-propagating flux, denoted by the superscript $+$; the corresponding backward-propagating behavior is shown in Fig. 7.

The diversity of modal flux among transmission eigenchannels arises from three factors: the directional flux-matrix elements $|t_{ms}^{+}(z)|^2$, whose magnitudes depend on the modal group velocity $\mathrm{v}_m$; the incident-eigenchannel modal weights $|v_{sn}|^2$; and modal alignment $\kappa_{mn}^{+}(z)$. The smaller diagonal elements for slower modes in $|t_{ms}^{+}(z)|^2$ [Fig. 3(a)] arise from the shorter distance these modes travel during one mean free scattering time, $\ell/\mathrm{v}_m$. This behavior is consistent with the reduced optical transmission observed for larger angles of incidence in disordered slabs, where smaller longitudinal velocity components lead to shorter penetration into the medium [45]. The mean free path is determined in simulations from the decay of coherence with the incident mode as a function of depth [53]; details are given in the Supplemental Material [46]. The off-diagonal elements increase over approximately one mean free path before decaying together with the diagonal elements [Fig. 3(a)]. Figure 3(b) shows the modal weights $|v_{sn}|^2$, which give the contributions of incident mode $s$to eigenchannel $n$.

The distribution of directional flux among the waveguide modes determines their associated forward- and backward-propagating eigenchannel fluxes $F_n^{\pm}(z)$ and their associated energy densities $u_n^{\pm}(z)$. Their ratio defines the corresponding directional eigenchannel velocity, as discussed in the Supplemental Material [46]:

$$\mathrm{v}_n^{\pm}(z) = \frac{F_n^{\pm}(z)}{u_n^{\pm}(z)}. \tag{4}$$

Here

$$F_n^{\pm}(z) = \sum_{m=1}^{N} |f_{mn}^{\pm}(z)|^2 \,, \tag{5}$$

and

$$u_n^{\pm}(z) = \sum_{m=1}^{N} |u_{mn}^{\pm}(z)|^2 = \sum_{m=1}^{N} \frac{1}{\mathrm{v}_m} |f_{mn}^{\pm}(z)|^2. \tag{6}$$

Here, $|u_{mn}^{\pm}(z)|^2$ is the energy density associated with the $m$th waveguide mode of the forward- or backward-propagating component of eigenchannel $n$. As the transmission-eigenchannel index $n$ increases, the incident modal weights $|v_{sn}|^2$ shift toward lower-velocity modes, leading to a decrease in the eigenchannel velocity $\mathrm{v}_n$ [45].

The incoherent sums of flux contributions from the incident modes, $\sum_{s=1}^{N=8} \mid t_{ms}^{+}(z) v_{sn} \mid^2$, with $N = 8$, are shown in Fig. 3(c) for transmission eigenchannels with $n = 1,2,3,8$. Beyond one mean free path, $\ell = 133a$, these sums decrease with depth and are smaller for lower velocity modes. The forward- and backward-propagating modal flux amplitudes within each eigenchannel,

$$f_{mn}^{\pm}(z) = \sum_{s=1}^{N=8} t_{ms}^{\pm}(z) v_{sn}\,, \tag{7}$$

are coherent sums weighted by the incident modal amplitudes at the input [Fig. 3(b)]. For transmission eigenchannels with higher-than-average transmission ($\tau_n > g/N$), the average forward modal flux, $|f_{mn}^{+}(z)|^2$, initially increases with depth for all modes [Fig. 3(d)]. Within these eigenchannels, the flux is larger for modes with higher modal group velocity $\mathrm{v}_m$ [Figs. 3(d) and 3(e)]. By contrast, for transmission eigenchannels with lower-than-average transmission ($\tau_n < g/N$), the modal flux decreases with depth [Fig. 3(d)] and is larger for lower-velocity modes $\mathrm{v}_m$ [Figs. 3(d) and 3(e)].

The ratio of coherent to incoherent flux for the forward-propagating component, $\kappa_{mn}^{+}$, which quantifies modal alignment, increases with depth for eigenchannels with $\tau_n > g/N$ and decreases for those with $\tau_n < g/N$ [Fig. 3(f)]. In the highest-transmission eigenchannel, $\kappa_{m1}^{+}(L)$ exceeds $70\%$ of its maximum possible value, where, more generally,

$$\kappa_{mn_{\max,L}}^{+} = \frac{(\sum_{s=1}^{N=8} |t_{ms}^{+}(L) v_{sn}|)^2}{\sum_{s=1}^{N=8} |t_{ms}^{+}(L) v_{sn}|^2}\,, \tag{8}$$

This maximum corresponds to perfect constructive interference when all complex modal contributions are in phase. The degree of destructive interference is greatest near transmission zeros [13,14] (Figs. 2(c)–(e)). Together, the panels of Fig. 3 show that the forward-propagating modal flux at depth $z$, $|f_{mn}^{+}(z)|^2$, is determined by three factors: the squared directional flux-matrix elements $|t_{ms}^{+}(z)|^2$, the input modal weights $|v_{sn}|^2$, and the alignment of the resulting modal contributions, quantified by $\kappa_{mn}^{+}(z)$.

Figure 4(a) shows simulations of the forward-propagating modal alignment, $\kappa^{+}_{mn}(z, L)$, as a function of depth for $N = 8$ and three sample lengths $L$. For transmission eigenchannels with above-average transmission, $\tau_n > g/N$, the modal alignment at fixed depth increases with sample length, reflecting stronger constructive interference in longer samples. By contrast, for eigenchannels with below-average transmission, $\tau_n < g/N$, the modal alignment at fixed depth decreases with sample length, reflecting stronger destructive interference. These trends are most pronounced for the fastest modes.

The output modal alignment of the fastest mode in the highest-transmission eigenchannel, $\kappa^{+}_{11}(L)$, increases monotonically with sample length and passes smoothly through the crossover to localization at $g = 1$, indicated by the dashed vertical line in Fig. 4(b). At $L = 2000a$, where $g = 0.43$, $\kappa^{+}_{11}(L)$ reaches 84% of its maximum possible value of 4.79. Thus, modal alignment approaches saturation before the system is deeply localized.

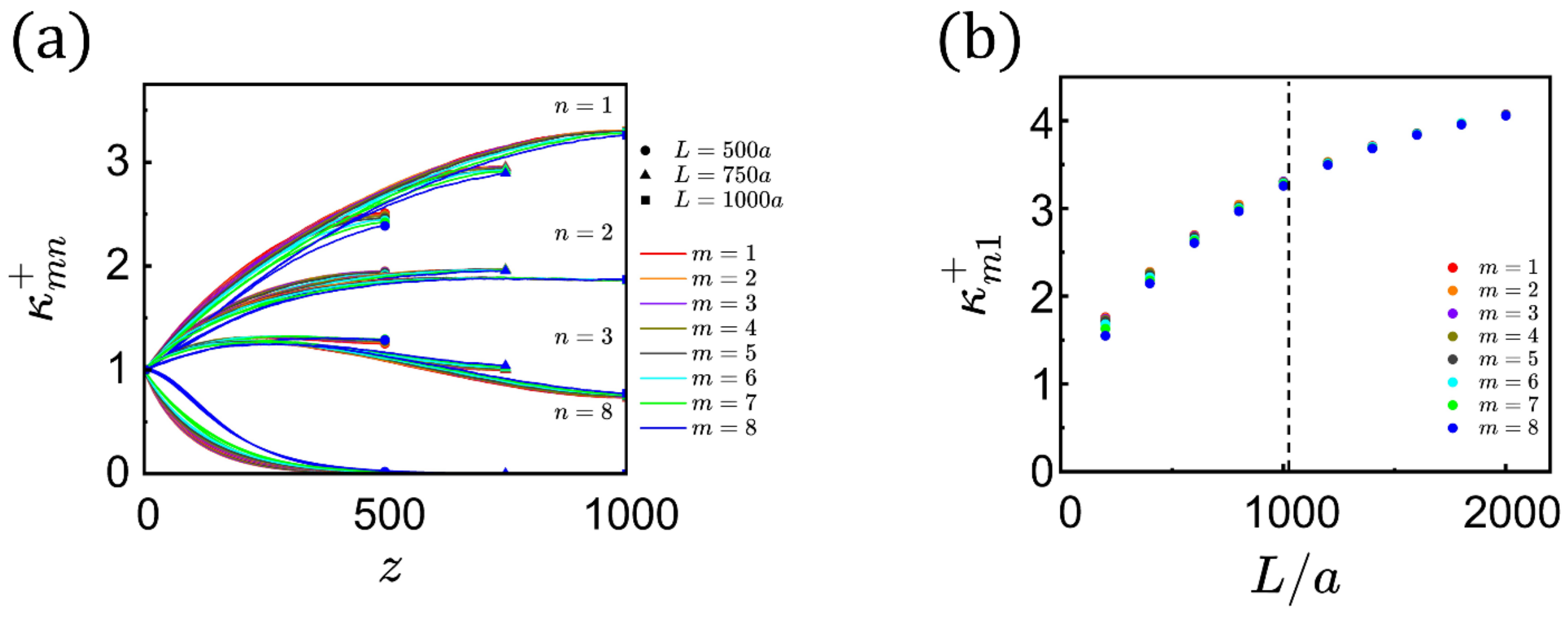

FIG. 4. Evolution of modal alignment with depth and with sample length. Simulations of $10^5$ random samples with $W = 26a$ and $\Delta\varepsilon = 0.3$. (a) Modal alignment $\kappa_{mn}^{+}$ versus depth $z$ for all modes $m$ contributing to eigenchannels $n = 1, 2, 3$, and 8, for sample lengths $L = 500a$ (circles), $750a$ (triangles), and $1000a$ (squares). (b) Modal alignment at the output surface in the highest-transmission eigenchannel, $\kappa_{m1}^{+}$, as a function of sample length $L$. Modal alignment for all modal components converges towards a common value as $L$ increases. The dashed vertical line marks the crossover to the localized regime $g < 1$.

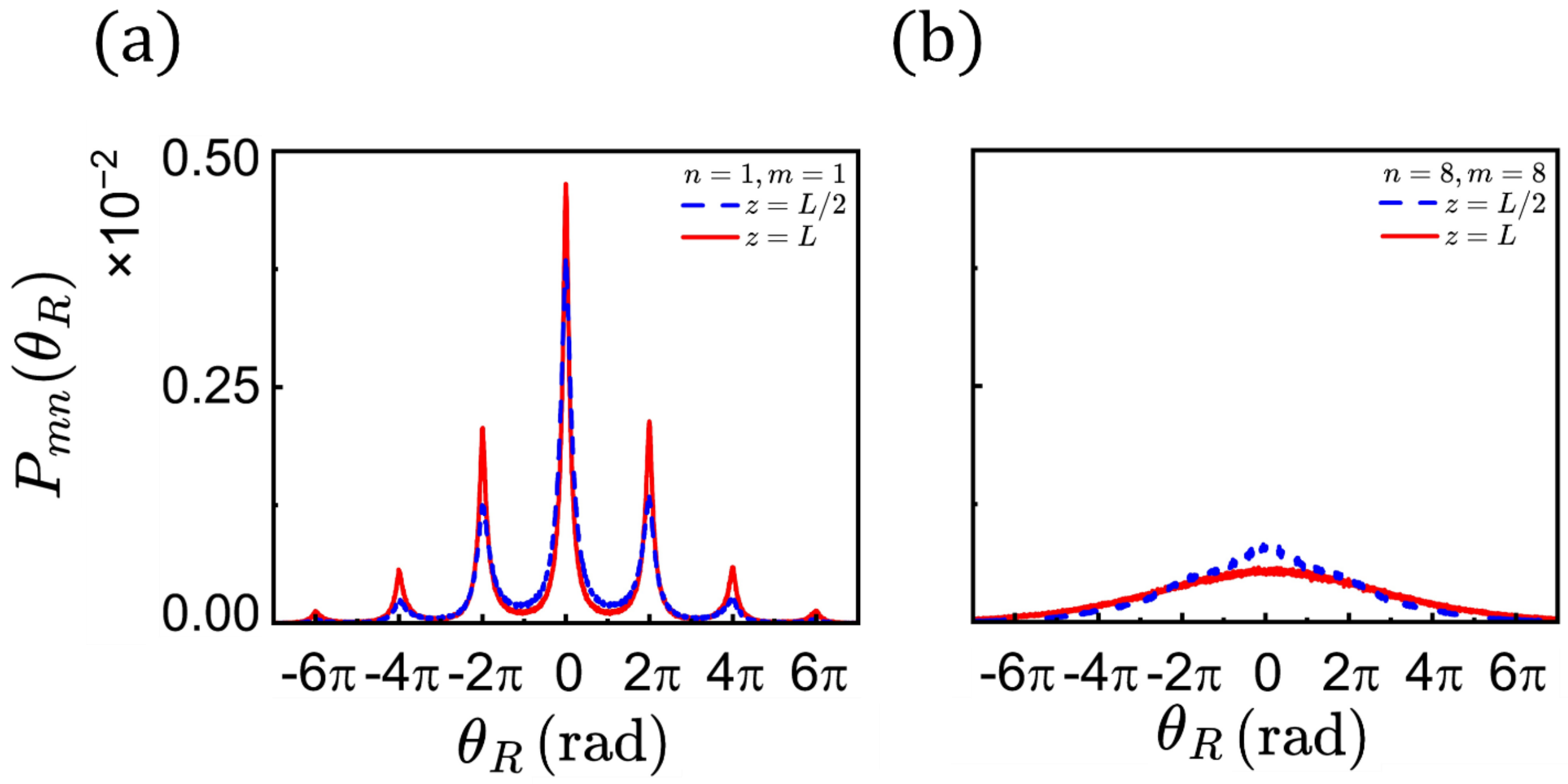

FIG. 5. Probability densities of angular distributions of flux amplitudes relative to the resultant in the highest- and lowest-transmission eigenchannels. Simulations of $3.12 \times 10^5$ random samples performed under the same conditions as in Fig. 4. (a) Probability density functions of the cumulative phases of flux amplitudes from the eight incident modes contributing to the $m = 1$ mode of the $n = 1$ transmission eigenchannel, computed relative to the resultant phase in each configuration, at depths L/2 and $L$. With increasing depth, the flux amplitudes increasingly align

with the resultant. (b) By contrast**,** the distribution of the cumulative phases of the flux components in the $n = 8, m = 8$ mode broadens with depth.

Modal alignment is further revealed by the probability density $P(\theta_R)$ of the cumulative phases of the flux-amplitude contributions $t_{ms}^{+}(z)v_{sn}$, measured relative to the resultant phase in each configuration (Fig. 5). For $m = n = 1$, $P(\theta_R)$ sharpens around $\theta_R = 0$ with increasing depth, indicating increasing alignment of the contributions [Fig. 5(a)]. The cumulative phase of each contribution is obtained by integrating its incremental phase changes with depth in the laboratory frame, without modulo-$2\pi$ wrapping. At each depth, the resultant is formed from the same contributions, and $\theta_R$is measured relative to its phase. Secondary peaks at $\theta_R = 2k\pi$, where $k$ is a nonzero integer, reflect the tendency of flux-amplitude vectors to return to alignment after passing through the opposite orientation at shallower depths. By contrast, for $m = n = 8$, the distribution broadens with depth as interference becomes increasingly destructive [Fig. 5(b)].

**E. Phase shifts across zeros and poles**

In the lowest-transmission eigenchannel, the phases of the individual incident-mode contributions $t_{ms}v_{sN}$to the $m$th output mode and of their resultant, $f_{mN} = \sum_{s=1}^{N} t_{ms}\, v_{sN}$, vary continuously with frequency except for abrupt jumps upon crossing a transmission zero [Figs. 2(c) and 2(d)]. In the experiment, these jumps are close to 180°, with the deviations arising primarily from the finite frequency step. They may also be reduced when absorption shifts the transmission zero below the real-frequency axis. In simulations, the jumps approach 180°as the frequency step is reduced. Here, both the contributing phasors and their resultant change discontinuously. This contrasts with a phase singularity in a speckle pattern, where the resultant field undergoes a $180°$ phase shift while the constituent phasors evolve continuously [54–56].

To identify the origin of these phase jumps, we examine the phase map of $\det(t)$ in the complex-frequency plane, shown in the bottom panel of Fig. 6(a). The imaginary frequency coordinate represents the field growth or decay rate $\gamma$. We obtain the map away from the real-frequency axis by computing $\arg\{\det(t)\}$ for a sample with uniform gain or loss, introduced through an imaginary component of the refractive index, $n_i = n_r\gamma/\omega$, where $n_r$ is the randomly selected real part of the refractive index and $\omega$is the angular frequency. Phase singularities identify the poles and zeros of $\det(t)$, including the real-axis transmission zeros near which the lowest transmission eigenvalue drops sharply, as shown in the top panel of Fig. 6(a).

We next examine which of the two factors forming each modal contribution, $t_{ms}v_{sN}$, produces the phase jump across a transmission zero: the transmission-matrix element $t_{ms}$ or the incident eigenchannel coefficient $v_{sN}$. Figure 6(b) shows the phase difference between adjacent frequency points for $v_{1N}$, which jumps discontinuously at each transmission zero; corresponding jumps occur in $v_{sN}$ for all input modes $s$. By contrast, the adjacent-frequency phase difference of the transmission-matrix element $t_{N1}$, shown in Fig. 6(c), varies rapidly only near poles, as illustrated for one pole in the top panel. Similar variations occur in $t_{ms}$ for all input modes $s$and output modes $m$. These results show that the discontinuous phase shifts of the modal contributions across transmission zeros originate primarily in phase jumps of the incident eigenchannel coefficients $v_{sN}$.

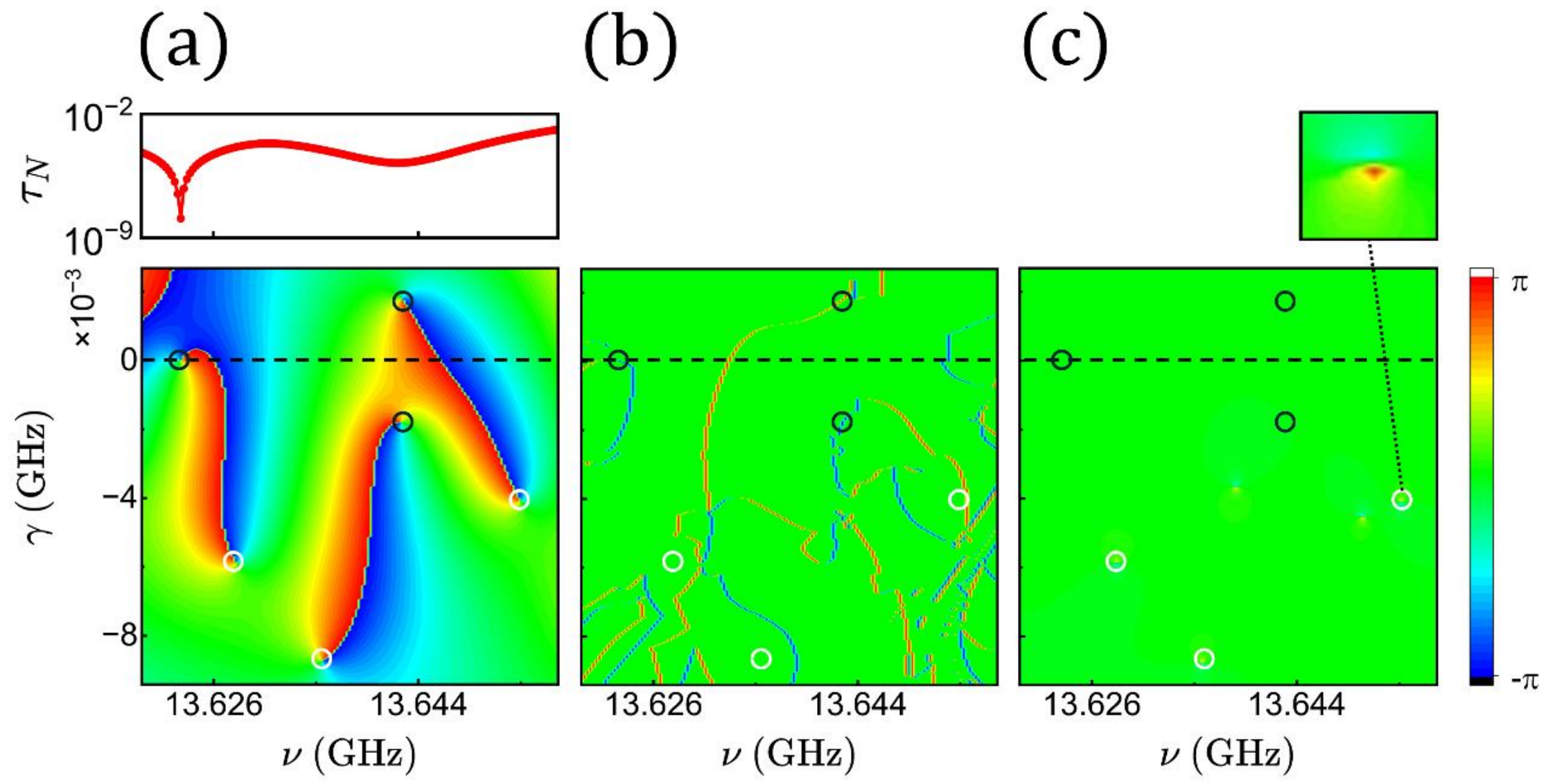

FIG. 6. Origins of transmission zeros and poles. Simulation of a random configuration with $W = 26a$, $L = 300a$, $\Delta\varepsilon = 0.3$, $g = 2.8$, and $N = 7$. (a) Phase map of $\arg\{\det(t)\}$ in the complex frequency plane. Black circles mark transmission zeros on the real axis ($\gamma = 0$, dashed line) or conjugate pairs of zeros, while white circles mark poles in the lower half-plane. (b) Map of the phase difference between adjacent frequencies for $v_{1N}$, the amplitude of input mode 1 in the lowest-transmission eigenchannel. Transmission zeros correspond to $\pi$ phase shifts in the waveform of the lowest transmission eigenchannel; poles produce no systematic phase shift. (c) Phase-difference map of the transmission-matrix element $t_{N1}$. Poles are accompanied by phase shifts totaling $\pi$ over a small range of $\gamma$, present for any input mode $s$ or output mode $m$; zeros produce no systematic phase shift.

### F. Eigenchannel flux, energy density, and velocity

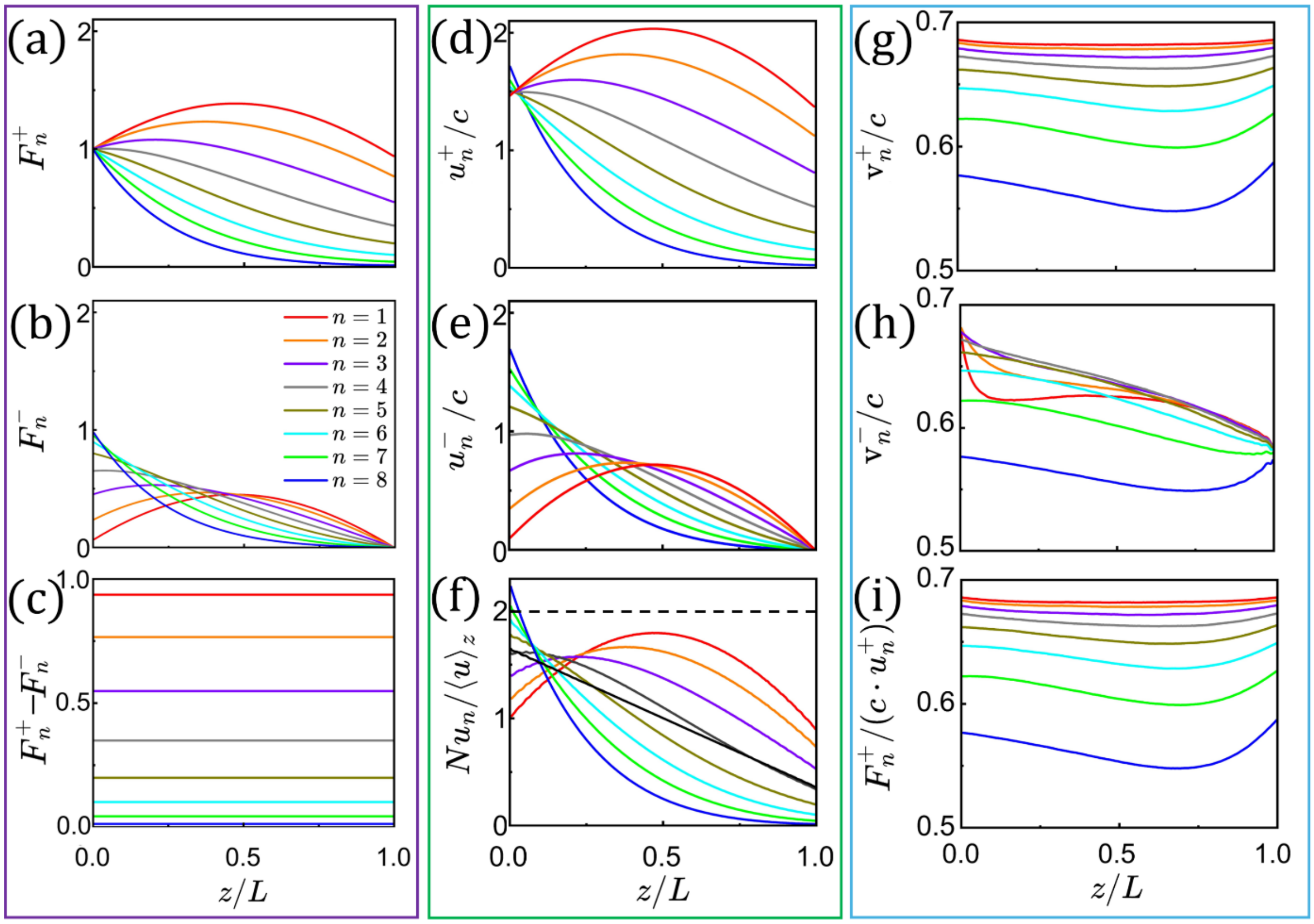

FIG. 7. Eigenchannel flux, energy density, and velocity. Simulations were performed for an ensemble of $5.5 \times 10^4$ random samples with the same disorder and dimensions as in Fig. 6. (a),(b); (d),(e); (g),(h) Forward- and backward-propagating fluxes, energy densities, and velocities for all transmission eigenchannels. The net flux, given by the difference between forward- and backward-propagating fluxes, is constant throughout the sample (c). The ensemble-averaged sum of counter-propagating energy densities is constant for excitation from both sides (dashed black line) and decreases linearly with depth for excitation from the left (solid black line) (f). Eigenchannel velocities $v_n^+(z)$ exhibit a minimum within the sample and return to their input values at the output surface (g),(h). (i) The ratio $F_n^+(z)/u_n^+(z)$ equals $v_n^+(z)$ [Eq. 4].

The forward- and backward-propagating fluxes in each eigenchannel are obtained by summing the corresponding modal fluxes, as given in Eq. (5). At the output surface, $F_n^+(L) = \tau_n$ and $F_n^-(L) = 0$. Figure 7 shows the forward- and backward-propagating fluxes, energy densities, and velocities for all eigenchannels in samples with the same disorder strength and width as those in Figs. 4–6, but with a shorter length, $L = 300a$, and higher conductance, $g = 2.96$. The net flux in each transmission eigenchannel, $F_n^+(z) - F_n^-(z)$, is independent of depth and equals $\tau_n$ [Fig. 7(c)]. By contrast, the average energy density decreases linearly with depth, as expected for diffusive transport and consistent with Fick's first law. The sum of the energy densities excited by unit-flux incidence in a transmission eigenchannel from the two sides of the sample is proportional to the density of states and is independent of depth [dashed black line in Fig. 7(f)].

The directional eigenchannel velocities $\mathrm{v}_n^{\pm}(z)$, which are flux-weighted harmonic means of the waveguide-mode velocities, have the same values at the input and output surfaces of the sample [38]. Both $\mathrm{v}_n^+(z)$ and $\mathrm{v}_n^-(z)$ decrease to minima within each eigenchannel before increasing toward the output surface and recovering their input-surface values [Figs. 7(g) and 7(h)]. The ratio $F_n^+(z)/u_n^+(z)$, plotted in Fig. 7(i), reproduces $\mathrm{v}_n^+(z)$ in Fig. 7(g). The depth dependence of $\mathrm{v}_n^{\pm}(z)$ compensates for that of the associated energy densities, while the net flux $F_n^+(z) - F_n^-(z) = \tau_n$ remains constant.

## 4. CONCLUSION

Coherence underlies established transmission eigenvalue scaling and statistics, including the bimodal distribution in diffusive systems, as well as Anderson localization. Here, microwave transmission matrix measurements and flux matrix simulations provide a direct description of the internal modal construction of transmission eigenchannels. We determine how the flux amplitude in each waveguide mode is formed from contributions seeded by all incident modes and how the magnitudes and phases of these contributions are organized throughout the sample. We show that the spatial evolution of modal flux within each transmission eigenchannel, together with the associated energy density and velocity, is governed by three factors: the squared directional flux-matrix elements, the incident-eigenchannel modal weights, and modal alignment. Modal alignment becomes strongly constructive in the highest-transmission eigenchannels, whereas near-complete destructive interference can produce vanishing transmission in the lowest-transmission eigenchannel at a transmission zero. The consistency of this description is demonstrated by the depth-independent net flux in every eigenchannel, given by the difference between its forward- and backward-propagating fluxes and equal to its transmission eigenvalue. Even near a transmission zero, the lowest transmission eigenvalue can be resolved many orders of magnitude below the noise floor of conventional transmission measurements because the individual modal contributions remain readily measurable as their coherent sum approaches zero. This capability may enable the coalescence of two transmission zeros at a square-root singularity to be measured [13] and used to sense minute changes in a sample.

One striking feature is the persistence of the widely different energy-density profiles of transmission eigenchannels, which depend on mutual coherence among many modal contributions, throughout samples much longer than the scattering mean free path over which individual modal phases are randomized. Our results show that this persistence is sustained by the systematic organization of modal alignment with depth. This depth-dependent organization provides a framework for understanding and controlling wave transport through complex systems and raises unique questions about how it is established and maintained within disordered media.

**ACKNOWLEDGMENTS**

The authors thank N. Koshy for valuable discussions. This work is supported by the National Science Foundation (US) under NSF-BSF Award No. 2211646 (AZG).

**DATA AVAILABILITY**

The data used in the current study are available from the corresponding author on reasonable request.